\documentstyle[eqsecnum,preprint,aps,prd]{revtex} 
\tighten
\begin{document}
\draft
\def\sqr#1#2{{\vcenter{\hrule height.3pt
      \hbox{\vrule width.3pt height#2pt  \kern#1pt
         \vrule width.3pt}  \hrule height.3pt}}}
\def\square{\mathchoice{\sqr67\,}{\sqr67\,}\sqr{3}{3.5}\sqr{3}{3.5}}
\def\today{\ifcase\month\or
  January\or February\or March\or April\or May\or June\or
  July\or August\or September\or October\or November\or December\fi
  \space\number\day, \number\year}
\def\oursh{{\rm sh}}
\def\ourch{{\rm ch}}
\def\thermal{thermal }
\def\csch{{\rm csch}}
\def\arsh{{\rm arsh}}
\preprint{WISC-MILW-94-TH-25} 
\title{Maximally Symmetric Spin-Two Bitensors on $S^3$ and $H^3$}
\author{Bruce Allen}
\address{
Department of Physics, University of Wisconsin -- Milwaukee\\
P.O. Box 413, Milwaukee, Wisconsin 53201, U.S.A.\\
email: ballen@dirac.phys.uwm.edu}
\date{\today}
\maketitle
\begin{abstract}
The transverse traceless spin-two tensor harmonics on $S^3$ and $H^3$
may be denoted by $T^{(kl)}{}_{ab}$.  The index $k$ labels the
(degenerate) eigenvalues of the Laplacian $\square$ and $l$ the other
indices.  We compute the bitensor $\sum_l T^{(kl)}{}_{ab}(x)
T^{(kl)}{}_{a'b'}(x')^*$ where $x,x'$ are distinct points on a sphere
or hyperboloid of unit radius.  These quantities may be used to find
the correlation function of a stochastic background of gravitational
waves in spatially open or closed Friedman-Robertson-Walker
cosmologies.
\end{abstract}
\vskip 0.3in

\section{INTRODUCTION}

It is generally accepted that our universe may be accurately modeled by
a Friedman-Robertson-Walker (FRW) cosmology \cite{HawkingEllis}.  In
FRW models, the spatial sections are maximally-symmetric Riemannian
three-manifolds.  In the spatially-flat FRW models, this is $R^3$ and
in the spatially-closed and spatially-open cases it is the three-sphere
$S^3$ and the three-hyperboloid $H^3$ respectively.

To study certain physical processes, such as the production of
gravitons in the early universe, it is useful to introduce a complete
set of transverse traceless symmetric rank-two tensors on this
three-space \cite{FordParker}.  These tensors may be denoted by
\begin{equation}
T^{(kl)}{}_{ab}(x)
\end{equation}
where $a$ and $b$ are the tensor indices, $k$ labels the eigenvalue,
and $l$ denotes the remaining degenerate index labels.  Explicit
formulae for the components of these tensors on $S^3$ and $H^3$ may be
found in work of Higuchi \cite{Higuchi} and Tomita \cite{Tomita}.  The
spectral functions of arbitrary-spin transverse traceless symmetric
tensors on $H^n$ and $S^n$ have been computed by Camporesi and Higuchi
\cite{CamporesiHiguchi}.

This note derives a simple, coordinate invariant, explicit expression
for the bitensor
\begin{equation}
W^{(k)}{}_{aba'b'}(x,x')= \sum_l T^{(kl)}{}_{ab}(x)
T^{(kl)}{}_{a'b'}(x')^*.
\end{equation}
Because the r.h.s. is a uniform sum over all the modes of a given
eigenvalue, it defines a maximally-symmetric bitensor, in the language
of Allen and Jacobson \cite{AllenJacobson}. 

To understand the significance of this bitensor, the reader might wish
to think about it's well-known scalar analogs.  In similar notation,
the sum over scalar harmonics on a unit two-sphere is
\begin{equation}
\sum_{l=-k}^k Y_{kl}(\Omega) Y_{kl}(\Omega')^* = 
{2 k + 1 \over 4 \pi} {\rm P}_k(\cos \mu),
\end{equation}
where the P's are Legendre polynomials and $\mu$ is the geodesic
distance between the points $\Omega$ and $\Omega'$ on the sphere.  A
less familiar example is the sum over the normalized scalar harmonics
on the unit three-hyperboloid,
\begin{equation}
\sum_l Y^{(kl)}(x) Y^{(kl)}(x')^* = {k \sin k \mu \over 2 \pi^2 \oursh \mu}.
\end{equation}
Further details concerning this identity may be found in
\cite{BanderItzykson}.

The bitensor $W^{(k)}{}_{aba'b'}$ calculated in this paper is the
spin-two equivalent of these quantities. It is useful in computing
Green functions (or two-point correlation functions, or propagators)
for gravitons (linearized gravitational excitations) in spatially open or
closed FRW models.  These Green functions have been determined in the
spatially flat case by Allen \cite{AllenGreen} and by Caldwell
\cite{CaldwellGreen} building on early work by Ford and Parker
\cite{FordParker}.  Similar methods have been used on a four-sphere to
find geometric expressions for the graviton propagator in de Sitter
space-time \cite{AllenTuryn}.

A brief outline of this paper follows.  In section \ref{secdef} we
define the tensor mode functions and their normalization.  The bitensor
is defined in \ref{secbit}, and expressed in terms of a fundamental set
of maximally-symmetric bitensors defined by Allen and Jacobson
\cite{AllenJacobson}.  The problem of finding the bitensor is then
reduced to finding three scalar functions.  We obtain a coupled set of
five differential equations which they obey.  In the section
\ref{solutions} these equations are solved in terms of hypergeometric
functions.  There remains an overall normalization constant, which is
determined in section \ref{findq}.  Finally, in a pair of appendices,
explicit expressions are given for the bitensor in terms of elementary
functions.  Appendix \ref{appendixH3} contains formulae for the
unit three-hyperboloid and appendix \ref{appendixS3} contains the
corresponding expressions for the unit three-sphere.

\section{DEFINITION AND NORMALIZATION OF MODES}
\label{secdef}
We work on the unit sphere $S^3$ and the unit hyperboloid $H^3$ with
metrics
\begin{equation}
ds^2 = 
\cases{
d\chi^2 + \sin^2 \chi (d\theta^2 + \sin^2 \theta d\phi^2) & 
with $\chi \in [0,\pi]$ for $S^3$, \cr \cr
d\chi^2 + \oursh^2 \chi (d\theta^2 + \sin^2 \theta d\phi^2) &
with $\chi \in [0,\infty)$ for $H^3$. \cr}
\end{equation}
The coordinates $\theta \in [0,\pi]$ and $\phi \in [0, 2\pi)$ range
over the unit two-sphere.  

Frequent use will be made of the geodesic distance $\mu(x,x')$ between
two points.  In terms of the coordinates of the two points this
distance is
\begin{eqnarray}
\rm{On \> S^3} \quad
\cos \mu &=& \cos \chi \cos \chi' + \sin \chi \sin \chi' \cos \gamma 
\qquad \mu \in [0,\pi],
\cr\cr
\rm{On \> H^3} \quad \quad
\ourch \mu & =& \ourch \chi \ourch \chi' - 
\oursh \chi \oursh \chi' \cos \gamma,
\qquad \qquad \mu \in [0,\infty).
\end{eqnarray}
On both the sphere and the hyperboloid 
\begin{equation}
\cos \gamma = 
\cos \theta \cos \theta' + \sin \theta \sin \theta' \cos (\phi-\phi')
\end{equation}
is the (cosine of the) angle between the points projected onto the
two-sphere

The mode functions are tensor eigenmodes of
the Laplace operator:
\begin{equation}
\square T^{(kl)}{}_{ab}(x) = \lambda_k T^{(kl)}{}_{ab}(x)
\end{equation}
which are transverse ($\nabla^a T^{(kl)}{}_{ab} =0$) traceless ($
T^{(kl)}{}_a{}^a =0$) and symmetric ($T^{(kl)}{}_{ab} =
T^{(kl)}{}_{ba}$).  The index $l$ is a multi-index denoting the
remaining degenerate labels.  Note that although the tensor harmonics
may be chosen to be real, we assume here that they are complex and use
$*$ to denote complex conjugation.

The index $k$ which labels the eigenvalues is discrete on $S^3$ and
continuous on $H^3$.  The eigenvalues are given by
\begin{equation}
\lambda_k = 
\cases{ 
-k^2 - 2 k + 2 & for $k=2,3,4,\cdots$ on $S^3$ \cr
-k^2 - 3 & for $k \in [0,\infty)$ on $H^3$. \cr}
\end{equation}
Note that the largest eigenvalue on $H^3$ is less than zero, as is also
the case for the eigenvalues of $\square$ acting on scalar functions.
The properties of the scalar eigenfunctions are explored in detail by
Bander and Itzykson \cite{BanderItzykson}, and their group theoretic
properties have been studied by  Vilenkin and Smorodinskii
\cite{NVilenkin}.

The multi-index $l$ refers to a set $\{ p,n,m\}$ of three discrete
indices.  The index $p$ is a polarization or parity index that takes
two possible values, $s$ or $v$.  The indices $n$ and $m$ label
harmonic functions $Y_{nm}(\theta,\phi)$ on the two-sphere.  The range
differs for $S^3$ and $H^3$:
\begin{eqnarray}
{\rm On } \> S^3 \qquad l &=& \cases{
p \in \{s,v\} & polarization\cr
n = 2,3,\cdots,k-1,k & two-sphere label \cr
m = -n,-n+1,\cdots,n-1,n & two-sphere label,\cr} \cr \cr \cr
{\rm On } \> H^3 \qquad  l &=& \cases{
p \in \{s,v\} & polarization\cr
n = 2,3,4,\cdots & two-sphere label \cr
m = -n,-n+1,\cdots,n-1,n & two-sphere label.\cr}
\end{eqnarray}
Note that the index $n$ starts at two because a spin-two field has no
monopole or dipole components.  We use the index $l$ collectively to
denote all three of these discrete indices.  Thus
\begin{eqnarray}
\label{multi}
{\rm On } \> S^3 \qquad
\delta_{ll'} \equiv \delta_{pp'} \delta_{nn'} \delta_{mm'}
\quad& {\rm and} & \quad \sum_l \equiv 
\sum_{p=s,v} \sum_{n=2}^k \sum_{m=-n}^n,\cr
{\rm On } \> H^3 \qquad
\delta_{ll'} \equiv \delta_{pp'} \delta_{nn'} \delta_{mm'}
\quad& {\rm and} & \quad \sum_l \equiv 
\sum_{p=s,v} \sum_{n=2}^\infty \sum_{m=-n}^n.
\end{eqnarray}
On the three-sphere $S^3$ the contracted $\sum_l 
\delta_l{}^l= 2(k-1)(k+3)$ is
finite, but on $H^3$ the contracted $\sum_l \delta_l{}^l=\infty$.

The tensors are normalized by the condition
\begin{equation}
\int \sqrt{g} d^{\;3 } x \>  T^{(kl)}{}_{ab}(x) T^{(k'l')}{}^{ab}(x)^*
= \cases{ 
\delta_{kk'} \delta_{ll'} & on $S^3$ \cr
\delta(k-k') \delta_{ll'} & on $H^3$ \cr}
\label{norm}
\end{equation}
where the integral is over the entire manifold.  Explicit formulae for
the components of these tensors may be found in Higuchi \cite{Higuchi}
for the case of $S^3$ and in Tomita \cite{Tomita} for the case of
$H^3$.  One may prove that these tensors modes are complete, in the
sense that any square-integrable transverse traceless rank-two
symmetric tensor may be expressed as a linear combination of the set.

\section{THE BITENSOR}
\label{secbit}
The bitensor $W$ is defined by the uniform sum over all the modes with
a given eigenvalue:
\begin{equation}
W^{(k)}{}_{aba'b'}(x,x') = \sum_l T^{(kl)}{}_{ab}(x)
T^{(kl)}{}_{a'b'}(x')^*.
\end{equation}
Note that the indices $a,b$ lie in the tangent space over the point $x$
and the indices $a',b'$ lie in the tangent space over the point $x'$.
For this reason, these indices can be contracted only if $x=x'$.  The
sum (or integral) of this bitensor over all $k$ is a projection
operator onto the space of symmetric rank-two transverse traceless
tensors \cite{CamporesiHiguchi}.

The bitensor $W^{(k)}{}_{aba'b'}$ is a {\it maximally-symmetric
bitensor} as defined by Allen and Jacobson \cite{AllenJacobson}.  (From
this point on, we assume familiarity with this reference.) It can thus
be expressed as a linear combination of the five different fundamental
maximally-symmetric bitensors with the correct index symmetries:
\begin{eqnarray}
\nonumber
W^{(k)}{}_{aba'b'}(x,x') & = & w^{(k)}_1(\mu) g_{ab} g_{a'b'} +
w^{(k)}_2(\mu) \left[n_a g_{ba'} n_{b'} + n_b g_{aa'} n_{b'}
      +n_a g_{bb'} n_{a'} + n_b g_{ab'} n_{a'}\right]  \cr\cr
+ & & 
w^{(k)}_3(\mu) \left[g_{aa'} g_{bb'} + g_{ba'} g_{ab'}\right] +
     w^{(k)}_4(\mu) n_a n_b n_{a'} n_{b'}  \cr\cr
+& & 
w^{(k)}_5(\mu) \left[g_{ab} n_{a'} n_{b'} + n_a n_b g_{a'b'}\right].
\end{eqnarray}
The functions $w^{(k)}_i(\mu)$ depend only upon the geodesic distance
$\mu$ between the points $x$ and $x'$.  The bitensor $g_a{}^{b'}(x,x')$
is the linear map which parallel transports a vector from $x$ along the
geodesic to $x'$.  The bitensor $n_a(x,x')$ is a unit-length vector
tangent to the geodesic at $x$, pointing away from $x'$, and
$n_{a'}(x,x')$ is a unit-length vector tangent to the geodesic at $x'$,
pointing away from $x$.  Further details may be found in Allen and
Jacobson \cite{AllenJacobson}.

The tracelessness of $W^{(k)}{}_{aba'b'}$ imposes two constraints on
the functions $w_i$.  (From this point on, we no longer indicate the
dependence of the functions $w^{(k)}_i(\mu)$ upon either the eigenvalue
label $k$ or the geodesic distance $\mu$.)
\begin{eqnarray}
w_4 &=&   9 w_1+ 4 w_2 + 6 w_3
\cr \cr
w_5 & =& - 3 w_1 - 2 w_3.
\end{eqnarray}
Thus one may reduce the problem of finding the bitensor of interest to
that of finding three unknown functions $ w_1$, $w_2$, and $w_3$.  In
terms of this set the bitensor of interest is
\begin{eqnarray}
\label{three}
W^{(k)}{}_{aba'b'}(x,x') & = &  
w_1  \left[ g_{ab} - 3 n_a n_b \right] 
       \left[ g_{a'b'} - 3 n_{a'} n_{b'} \right]  \cr\cr
+& & w_2  \left[ g_{bb'} n_a n_{a'} + g_{ab'} n_b n_{a'} + 
                 g_{ba'} n_a n_{b'} + g_{aa'} n_b n_{b'} + 
                 4 n_a n_b n_{a'} n_{b'} \right]   \cr\cr
+& & w_3  
\left[ g_{ab'} g_{ba'} + g_{aa'} g_{bb'} - 2 g_{a'b'} n_a n_b - 
        2 g_{ab} n_{a'} n_{b'} + 6 n_a n_b n_{a'} n_{b'} \right].
\end{eqnarray}
This expression is traceless on either index pair $ab$ or $a'b'$.

The requirement that the bitensor be transverse $\nabla^a
W^{(k)}{}_{aba'b'}=0$ imposes a pair of additional constraints.  These
are first-order differential equations that must be obeyed by the
$w_i$.
\begin{eqnarray}
0 &=& {w'}_1 + 3 A w_1 + {w'}_3 + 3 A w_3 -C w_2 + C w_3 
\label{tran1}
\\ \nonumber \\
0 &=& {w'}_2 + 3 A w_2 - {w'}_3 - 3 A w_3 -3 C w_1 -5 C w_3.
\label{tran2}
\end{eqnarray}
Here $w' \equiv dw(\mu)/d\mu$.  The real functions $A(\mu)$ and
$C(\mu)$ arise from differentiating the fundamental bitensors $n_a$,
$g_{ab'}$ and $n_{b'}$.  They are given by Allen and Jacobson
\cite{AllenJacobson} as
\begin{equation}
A= {1 \over r} \cot (\mu/r) \quad {\rm and} \quad 
C= -{1 \over r} \csc (\mu /r),
\end{equation}
where $r=1$ for a unit-radius sphere $S^3$ and $r=i$ for a unit-radius
hyperboloid $H^3$.   The quantity $C^2 - A^2 = 1/r^2 = \pm 1$ with the
upper sign for $S^3$ and the lower sign for $H^3$.

The final equation obeyed by the bitensor is the eigenvalue condition
\begin{equation}
\left( \square - \lambda_k \right ) W^{(k)}{}_{aba'b'} =  0.
\end{equation}
When applied to the bitensor this yields three coupled second-order
differential equations, which are
\begin{eqnarray}
0 & = &  {w''}_1 + 2 A {w'}_1 - 
( 6 (A^2  +  C^2)  + \lambda_k ) w_1 - 4 (A^2  + C^2 ) w_3
\label{wave1}
\\ \nonumber \\
0 & = & {w''}_2 + 2 A {w'}_2 + 18 A C w_1 - 
(  5 A^2  - 4 A C + 5 C^2  + \lambda_k ) w_2 + 
3 (A^2  + 6 A C + C^2 ) w_3
\label{wave2}
\\ \nonumber \\
0 & = & {w''}_3 + 2 A {w'}_3 + 4 A C w_2 - 
      ( 2 A^2  + 4 A C + 2 C^2  +\lambda_k ) w_3
\label{wave3}
\end{eqnarray}
These equations are easily solved.

\section{Solutions of the transverse and eigenvalue equations}
\label{solutions}
To solve equations (\ref{tran1}-\ref{tran2}) and
(\ref{wave1}-\ref{wave3}) it is convenient to introduce a new variable
\begin{equation}
\alpha^{(k)}(\mu) = w^{(k)}_1(\mu) + w^{(k)}_3(\mu).
\end{equation}
As earlier, we drop $k$ and $\mu$ and denote this by $\alpha$.
Equation (\ref{tran1}) implies that $\alpha$ satisfies the first-order
equation
\begin{equation}
\alpha' + 3 A \alpha = C(w_2 - w_3).
\label{g1}
\end{equation}
The sum of (\ref{wave1}) and (\ref{wave3}) implies that $\alpha$
satisfies the second-order equation
\begin{equation}
\alpha '' + 2 A \alpha' - (6(A^2 + C^2) + \lambda_k) \alpha = 
-4 A C (w_2 - w_3).
\label{g2}
\end{equation}
Using (\ref{g1}) to replace the r.h.s. of (\ref{g2}), together with the
relation $C^2 - A^2 = 1/r^2$ yields
\begin{equation}
\alpha '' + 6 A \alpha' - \left[{6 \over r^2} + \lambda_k\right] \alpha
= 0.
\label{alphaeq}
\end{equation}
This has the same form as equation (2.3) of Allen and Jacobson
\cite{AllenJacobson}, and thus may be transformed into the
hypergeometric differential equation by defining a new variable $z =
\cos^2 (\mu/2r)$.

The hypergeometric equation has a pair of independent solutions.  One
of these solutions is singular at $z=1$, when the two points $x$ and
$x'$ are coincident, and $\mu=0$.  However $W^{(k)}{}_{aba'b'}$ is
finite everywhere, and this solution must be discarded.  The other
solution, which is regular at $\mu=0$, is given by
\begin{equation}
\alpha = Q_k \; {}_2 F_1(3 + \sqrt{3 - r^2 \lambda_k},
3 - \sqrt{3 - r^2 \lambda_k};7/2;1-z).
\end{equation}
where $Q_k$ is a normalization constant which is determined in section
\ref{findq} and ${}_2F_1$ is the Gauss hypergeometric function.  Since
${}_2F_1(a,b;c;z)$ is symmetric in $a$ and $b$, one may choose either
sign of the square root without changing $\alpha$.

The properties of the hypergeometric function differ considerably for
the sphere and the hyperboloid.  The constants $a$, $b$, and $c$ in the
two cases are given by
\begin{equation}
{\rm On} \> S^3 \cases{
a = 4 + k\cr
b = 2 - k \cr
c = 7/2 \cr} \qquad
{\rm On} \> H^3 \cases{
a = 3 + ik\cr
b = 3 - ik \cr
c = 7/2 \cr}.
\end{equation}
Thus on the sphere, where $b$ is a non-positive integer, the
hypergeometric function is a polynomial of order $k-2$ in $z$, whereas
on $H^3$ it is an associated Legendre function
\cite{GradshteynRyzhik}.  (The solutions are given explicitly
in Appendix \ref{appendixH3} for $H^3$ and in 
appendix \ref{appendixS3} for $S^3$.)

It is straightforward to solve the remaining equations.
From (\ref{g1}) and (\ref{tran2}) one has
\begin{eqnarray}
w_1 + w_3 &=& \alpha \cr
w_2 - w_3 &=& C^{-1} \left({d \over d\mu}  + 3 A \right) \alpha \cr
3 w_1 + 5 w_3 &=&  C^{-1} \left({d \over d\mu}  + 3 A \right)
 (w_2 - w_3).
\label{lineq}
\end{eqnarray}
The solution is easily written in terms of a pair of functions
\begin{equation}
\alpha(z) = Q_k \; {}_2F_1(a,b;c;1-z) \quad {\rm and} \quad
\beta(z) = Q_k \; {}_2F_1(a+1,b+1;c+1;1-z ).
\label{aandb}
\end{equation}
Inverting the linear equations (\ref{lineq}) one obtains
\begin{eqnarray}
w_1 &=&  \left[ 2(\lambda_k r^2 - 6) z(z-1) - 2 \right] \alpha(z) + 
{4 \over 7} \left[ (\lambda_k r^2 + 6) z (z - {1 \over 2}) (z-1) \right] 
\beta(z), \cr\cr
w_2 &=&  2 (1-z) \left[ (\lambda_k r^2 - 6) z + 3 \right] \alpha(z) - 
{4 \over 7} \left[ (\lambda_k r^2 + 6) z (z-1) (z - {3 \over 2}) \right]
 \beta(z), \cr\cr
w_3 &=&   \left[ -2 (\lambda_k r^2 - 6) z(z-1) + 3 \right] \alpha(z) -
{4 \over 7} \left[ (\lambda_k r^2 + 6) z (z - {1 \over 2}) (z-1) \right] 
\label{explicit}
\beta(z).
\end{eqnarray}
One may verify by direct substitution that these functions satisfy the
two transversality constraints (\ref{tran1}-\ref{tran2}) and the three
eigenvalue constraints (\ref{wave1}-\ref{wave3}).

\section{THE NORMALIZATION COEFFICIENT Q}
\label{findq}
There remains a single undetermined normalization constant $Q_k$.  To
determine it, consider the biscalar quantity
\begin{eqnarray}
g^{aa'} g^{bb'} W^{(k)}{}_{aba'b'}(x,x') &=& 6 w_1 - 4 w_2 + 14 w_3 \cr\cr
= \left[ -8 z(z-1) \lambda_k r^2 + 48 z (z-{1 \over 2}) + 6 \right]
\alpha(z) &- & {16 \over 7} \left[ (\lambda r^2 +6 ) z (z + {1 \over
2}) (z-1)\right]
 \beta(z).
\end{eqnarray}
In the coincident limit as $x' \to x$ and $z \to 1$
one obtains
\begin{equation}
\label{coin}
W^{(k)}{}_{ab}{}^{ab}(x,x) = \sum_l 
 T^{(kl)}{}_{ab}(x) T^{(kl)}{}^{ab}(x)^* =30  \alpha(1) =30 Q_k.
\end{equation}
At this point, one must proceed differently for the cases of $S^3$ and
$H^3$

To determine the normalization $Q_k$ for the three-sphere, integrate
both sides of (\ref{coin}) over the unit three-sphere (of volume $2
\pi^2$).  Set the index $l=l'$ and then sum over $l$, using
the normalization condition (\ref{norm}) 
and the contraction of the multi-index
given immediately after equation 
(\ref{multi}). Setting $k=k'$, one obtains the normalization
condition
\begin{equation}
{\rm On} \> S^3 : \quad 2 \pi^2 (30 Q_k ) = \sum_l 
\delta_l{}^l = 2 (k-1)(k+3)
\Rightarrow Q_k = {(k-1)(k+3) \over 30 \pi^2}.
\end{equation}
Together with equations (\ref{three}), (\ref{aandb}) and
(\ref{explicit}) this completely determines the bitensor of interest.

To determine the normalization $Q_k$ for the three-hyperboloid, one
must proceed slightly differently.  The previous technique fails
because setting $k=k'$ in the normalization condition (\ref{norm}) give
infinity on both sides, and the contraction $\sum_l 
\delta_l{}^l$ is also
infinite.  However Camporesi and Higuchi \cite{CamporesiHiguchi} have
calculated the spectral function in hyperbolic spaces.  In our
notation, this is defined by
\begin{equation}
\mu(k) = \pi^2 \sum_l T^{(kl)}{}_{ab}(x) T^{(kl)}{}^{ab}(x)^*.
\end{equation}
(Note that $\mu(k)$ is Camporesi and Higuchi's notation for the
spectral function, and {\it does not refer to the geodesic distance
$\mu$ as used in this paper.)} On $H^3$ they obtain $\mu(k) = k^2 +
4$.  Using (\ref{coin}) this implies that
\begin{equation}
{\rm On} \> H^3 : \quad \mu(k) = k^2 + 4 = 30 \pi^2  Q_k 
\Rightarrow Q_k = {k^2 + 4 \over 30 \pi^2}.
\end{equation}
Together with equations (\ref{three}), (\ref{aandb}) and
(\ref{explicit}) this completely determines the bitensor of interest.
We note that the $H^3$ normalization agrees with naive expectations;
one would expect that sending $k \to -1 + ik$ and $\chi \to i\chi$
performs the required
analytic continuation from $S^3$ to $H^3$.  This is indeed the case
here.

\acknowledgments
The author thanks Robert Caldwell and Atsushi Higuchi for a number of
useful conversations.  This work has been partially supported by
National Science Foundation grant number PHY91-05935.

\appendix
\section{Explicit formula for $H^3$}
\label{appendixH3}
In this section, explicit expressions for the coefficients $w_i(\mu)$
are found for the hyperbolic space $H^3$. The hypergeometric functions
appearing in (\ref{aandb}) may be expressed in terms of associated
Legendre functions via equation (15.4.18) of \cite{AbramowitzStegun}.
\begin{eqnarray}
\alpha(\mu) &=& 15 Q_k \sqrt{ \pi/2} (\oursh \mu)^{-5/2} 
P^{-5/2}_{-1/2 + ik}(\ourch \mu) \cr
\beta(\mu) &=& 15 Q_k \sqrt{ \pi/2} (\oursh \mu)^{-7/2} 
P^{-7/2}_{-1/2 + ik}(\ourch \mu)
\end{eqnarray}
These functions are the {\it slant P}'s of reference
\cite{GradshteynRyzhik} whose branch cuts lie to the left of $z=1$ on
the real axis.  There is a simple relationship between $\alpha$ and
$\beta$:
\begin{equation}
\beta = - 7{ \csch \mu \over k^2 + 9} {d \alpha \over d\mu}.
\end{equation}
The Legendre functions may be expressed in terms of elementary
functions:
\begin{eqnarray}
P^{-5/2}_{-1/2 + ik}(\ourch \mu) 
& = &
\sqrt{ 2 \over \pi \oursh\mu }(1+k^2)^{-1}(4+k^2)^{-1}
\Biggl[ -3 \cos k \mu \coth  \mu  \cr
&&
\qquad \qquad +{\sin k \mu \over 2 k} 
 \left(( 2 - k^2) ( 1 + \coth^2 \mu )  +
   ( 4 + k^2) \csch^2 \mu)  \right) \Biggr]\\
P^{-7/2}_{-1/2 + ik}(\ourch \mu) 
& = &
\sqrt{ 2 \over \pi \oursh\mu }
(1+k^2)^{-1}(4+k^2)^{-1}(9+k^2)^{-1}\Biggl[
(k^2 - 15 \csch^2 \mu - 11) \cos k \mu   \cr
&&
\qquad \qquad + 6 {\sin k \mu \over k} 
\left( (1-k^2) \coth^3 \mu + (k^2 + {3 \over 2})
\coth \mu \; \csch^2 \mu \right) \Biggr]. \nonumber
\end{eqnarray}
Making use of these expressions, the relation $2z=1+\ourch \mu$,
and the explicit expressions (\ref{explicit}) for the $w_i$ yields
\begin{eqnarray}
w^{(k)}_1(\mu) &=& {\csch^5 \mu \over 4 \pi^2 (k^2+1)}
\Biggl[  {\sin(k \mu) \over k} 
\left(3 + (k^2 + 4) \oursh^2 \mu - k^2 (k^2+1) \oursh^4 \mu \right) \cr
&& \qquad \qquad  \quad -    \cos(k \mu)
 \left(3 / 2 + (k^2+1) \oursh^2 \mu \right) \oursh 2 \mu \Biggr]\cr
w^{(k)}_2(\mu) &=& 
{\csch^5 \mu \over 4 \pi^2 (k^2+1) }
\Biggl[  {\sin(k \mu) \over k}
\left(3 + 12 \ourch \mu - 3 k^2 (2 \ourch \mu + 1) \oursh^2 \mu +
k^2 (k^2 +1 ) \oursh^4 \mu \right) \cr
&& \qquad \quad 
+    \cos(k \mu)  
 \left(-12 - 3 \ourch \mu + 2 (k^2-2) \oursh^2 \mu
+ 2 (k^2 +1) \ourch \mu \; \oursh^2 \mu \right) \oursh \mu \Biggr]\cr
w^{(k)}_3(\mu) &=& {\csch^5 \mu \over 4 \pi^2 (k^2+1)}
\Biggl[  {\sin(k \mu) \over k}
\left(3 - 3k^2 \oursh^2\mu + k^2(k^2+1) \oursh^4 \mu \right) \cr
&& \qquad \qquad  \quad +  \cos(k \mu)
 \left( - 3 / 2  + (k^2+1) \oursh^2 \mu  \right) \oursh 2 \mu \Biggr].
\end{eqnarray}
By direct substitution, one may verify that these functions satisfy 
the transverse conditions (\ref{tran1}-\ref{tran2}), the
eigenvalue equations (\ref{wave1}-\ref{wave3}), and the
normalization condition.

\subsection{Behavior near $\mu=0$}
These solutions are
regular in the neighborhood of $\mu = 0$, since
\begin{equation}
\lim_{\mu \to 0} \pmatrix{w_1 \cr w_2 \cr w_3} =
\left[ {k^2 + 4 \over 420 \pi^2} \right]
\pmatrix{
 -28 \>+ & (10 k^2 + 6 ) \mu^2 \cr
         & (-6 k^2 + 9 ) \mu^2 \cr
  42 \>- & (11 k^2 + 15) \mu^2} + {\rm O} (\mu^4).
\end{equation}
Since there is a {\it unique} solution to (\ref{alphaeq}) which is regular at
$\mu=0$ this proves that the functions given above are the 
{\it correct}
solutions to the transverse conditions and eigenvalue equations.

\subsection{Behavior as $k \to 0$}
It is remarkable that in the limit $k \to 0$ these functions approach
finite limits, rather than vanishing as one might naively expect.
\begin{equation}
\lim_{k \to 0} \pmatrix{w_1 \cr w_2 \cr w_3} =
{\csch^5 \mu \over 16 \pi^2}
\pmatrix{
4\mu + 8 \mu \ourch 2 \mu - 4 \oursh 2 \mu - \oursh 4 \mu \cr
12 \mu + 48 \mu \ourch \mu - 36 \oursh \mu 
- 8 \oursh 2 \mu - 4\oursh 3 \mu +
\oursh 4 \mu   \cr
12 \mu - 8 \oursh 2 \mu + \oursh 4 \mu }
+ {\rm O}(k^2)
\end{equation}
The corresponding biscalar identity for the scalar eigenfunctions of
the Laplacian $\square$ is
\begin{equation}
\sum_l Y^{(kl)}(x) Y^{(kl)}(x')^* = 
{k \sin k \mu \over 2 \pi^2 \oursh \mu}.
\end{equation}
In the scalar case, unlike the tensor case, the r.h.s. vanishes in the
small-$k$ limit.

\section{Explicit formula for $S^3$}
\label{appendixS3}
In this section, explicit expressions for the coefficients $w_i(\mu)$
are found for the three-sphere $S^3$.  The functions $\alpha$ and
$\beta$ may be expressed as power series of order $k-2$ in $\cos \mu$.
This is because the hypergeometric series terminate, since $b$ is zero
or a negative integer.  (The one exception occurs if $k=2$ for the
function $\beta$; but it is not needed because it's coefficient in the
expressions (\ref{explicit}) for the $w_i$ vanishes when $k=2$.)

The hypergeometric functions appearing in (\ref{aandb}) may be also
expressed in terms of Legendre polynomials of $\cos \mu$ via equation
(15.4.19) of \cite{AbramowitzStegun}.
\begin{eqnarray}
\alpha(\mu) &=& 15 Q_k \sqrt{ \pi/2} (\sin \mu)^{-5/2} 
{\rm P}^{-5/2}_{k+1/2}(\cos \mu) \cr
\beta(\mu) &=& 15 Q_k \sqrt{ \pi/2} (\sin \mu)^{-7/2} 
{\rm P}^{-7/2}_{k+1/2}(\cos \mu)
\end{eqnarray}
These functions are the {\it straight} P's of reference
\cite{GradshteynRyzhik} which have no branch cuts on the real axis
between $0 \le z \le 1$. There is a simple relationship between
$\alpha$ and $\beta$:
\begin{equation}
{d \alpha \over d\mu} = - {1 \over 7} (k-2)(k+4) \beta \sin \mu.
\end{equation}
One may calculate these Legendre polynomials using derivative formula,
the terminating hypergeometric series, or recursion relations.

While it is possible to give a series form for the $w_i$, it is not
very illuminating.  Rather, we illustrate the general behavior by
giving the forms of $w_i$ for the first three values of $k$:
\begin{eqnarray}
{\rm for} \> k=2 \quad
\pmatrix{w_1 \cr w_2 \cr w_3} &=& {1 \over 6 \pi^2}
      \pmatrix{ 4 - 6 \cos^2 \mu \cr
   -3 - 3 \cos \mu + 6 \cos^2 \mu \cr
   -3 + 6 \cos^2 \mu }
\cr\cr
{\rm for} \> k=3 \quad
\pmatrix{w_1 \cr w_2 \cr w_3} &=& {2 \over 5 \pi^2}
   \pmatrix{
    8 \cos \mu - 10 \cos^3 \mu \cr
    1 - 7 \cos \mu - 4 \cos^2 \mu + 10 \cos^3 \mu \cr
   -7 \cos \mu + 10 \cos^3 \mu }
\cr\cr
{\rm for} \> k=4 \quad
\pmatrix{w_1 \cr w_2 \cr w_3} &=& {1 \over 10 \pi^2}
   \pmatrix{
-12 + 118 \cos^2 \mu - 120 \cos^4 \mu \cr
11 + 19 \cos \mu - 100 \cos^2 \mu - 40 \cos^3 \mu + 120 \cos^4 \mu \cr
11 - 110 \cos^2 \mu + 120 \cos^4 \mu}
\end{eqnarray}
One may also obtain recursion relations for the $w_i$ which follow from
the corresponding recursion relations for the Legendre polynomials.


\end{document}